\documentstyle[hpgl,12pt]{article} 
\normalbaselines
\begin{document}
\bibliographystyle{unsrt}

\vbox {\vspace{6mm}} 

\begin{center}
{\large \bf PHOTON STATISTICS \\ OF A TWO-MODE SQUEEZED VACUUM} \\[7mm]
G. Schrade,
V. M. Akulin \footnote{also Moscow Institute of Physics and
Technology Dolgoprudny, Moscow, Russia},
W. P. Schleich \footnote{also Max-Planck-Institut f\"ur Quantenoptik,
D-85748 Garching, Germany}\\
{\it Abt. f\"ur Quantenphysik, Universit\"at Ulm \\
D-89069 Ulm, Germany }\\[5mm]
V. I. Man'ko \footnote{permanent address: Lebedev Institute of the
Academy of Sciences, Moscow, Russia}\\
{\it Dipartimento di Scienze Fisiche \\ Universita di Napoli Federico II\\
80125 Napoli, Italy}\\[5mm]
\end{center}

\vspace{2mm}

\begin{abstract}
We investigate the general case of the photon distribution of
a two-mode squeezed vacuum and show that
the distribution of photons among the two modes depends on four
parameters: two squeezing parameters, the relative phase between the
two oscillators and their spatial orientation.
The distribution of the total number of photons
depends only on the two squeezing parameters.
We derive analytical expressions
and present pictures for both distributions.
\end{abstract}

\section{Introduction}
Squeezing the quantum fluctuations of the radiation field has been
demonstrated experimentally using various optical systems \cite{squeeze}.
Most of them
rely on two-mode squeezing \cite{Caves1,Barnett,Yurke}.
Therefore the properties of a two-mode
squeezed state have been studied extensively \cite{Schumaker,Knight}.
However it appears, that
the photon statistics of such a state has not been investigated in
all details. Only some particular cases have been considered
(see for example Refs. \cite{Caves} and \cite{Artoni}).
In the present paper we therefore extend these considerations to an
arbitrary two-mode squeezed vacuum and
address the following questions:
(i) What is the most general case of the squeezed vacuum of a two-mode
oscillator,
and how many independent parameters are needed to describe this state?
(ii) What is the photon statistics in this state?
This is of interest in the context of the degenerate parametric
amplifier \cite{Singh} since the mathematical structure of the
two-mode distribution function coincides with the transition probability
function of a two-dimensional parametric oscillator \cite{Manko}.

\section{general case of the squeezed vacuum of a two-mode oscillator}
We start our considerations with the question (i).
The linear canonical transformation
\begin{equation}
   (a^\dagger,a;b^\dagger,b)\longrightarrow
   (A^\dagger,A;B^\dagger,B)
\label{eq:one}\end{equation}
of the creation operators $a^{\dagger}$,
$b^{\dagger}$ and the annihilation operators $a$, $b$ to their new
counterparts $A^\dagger$, $B^\dagger$ and $A$, $B$
corresponding to the two modes
suggests a set of 10 parameters.
Indeed, the generic transformation (\ref{eq:one})
of the four-dimensional vector
$(a^\dagger,a,b^\dagger,b)$ into $(A^\dagger,A,B^\dagger,B)$
via a $4\times4$ matrix brings in 16 complex values,
that is 32 parameters.
Since $a$ and $a^\dagger$, $b$ and $b^\dagger$
,$A$ and $A^\dagger$, $B$ and $B^\dagger$
are hermitian conjugate of each other, only half of these
parameters are independent.
Therefore the number of parameters reduces to 16.
The condition to preserve the commutation relation $[A,B]=0$,
$[A,B^\dagger]=0$, $[A,A^\dagger]=1$, and $[B,B^\dagger]=1$ provides
additional constraints.
Since the commutator of an operator with its hermitian conjugate
is always real, the last two conditions provide only two constraints,
whereas the first two conditions must hold for the real and
the imaginary part separately.
This decreases the number of parameters by six leaving us
indeed with 10 parameters.

This finding is in accordance with the results of group theory.
Reference \cite{Milburn}
shows, that the group of rotations and squeezing of the
four-dimensional phase space, that preserves the phase space volume
of the two degrees of freedom, consists of 10 generators.
Hence there are 10 parameters determining the elements of the
real symplectic group Sp(4,R).

But what is the physical meaning of these parameters?
They are associated with rotation and squeezing transformation
of phase space of two oscillators.
Generic rotations of four-dimensional phase space are described by
6 parameters but the symplectic rotations in four-dimensional space,
that is the transformation preserving the commutation relations, are
described by 4 parameters.
They can be represented as a
sequence of four rotations. The first rotation given by the
transformation
$e^{i \hat{M} \phi_1}$, where $\hat{M} = i ( a^\dagger b - b^\dagger a)$
is the angular momentum operator, corresponds to the rotation in the
coordinate space by the angle $\phi _1$.
The second and the third rotations are given by the operator
$e^{i \psi a^\dagger a}$ and $e^{i \chi b^\dagger b}$, and correspond
to rotations by the angles $\psi $ and $\chi $ in the
respective phase-spaces
of the two oscillators.
The last rotation again can be taken in the form
$e^{i \hat{M} \phi_2} $.
Thus the angles $\phi _1$, $\phi _2$, $\psi$ and $\chi$
are chosen
to be the parameters of the symplectic rotation.
The general symplectic transformation also includes squeezing.
We can
represent the generic real symplectic transformation (\ref{eq:one})
as consisting of three consecutive transformations:
symplectic rotation (4 parameters),
followed by independent
squeezing of the two modes given by the squeezing operators
\begin{equation}
   \hat{S_1} = e^{r_1(a^2-a^{\dagger ^2})/2}, \qquad
   \hat{S_2} = e^{r_2(b^2-b^{\dagger ^2})/2},
\label{squeeze}\end{equation}
(2 squeezing parameters $r_1$ and $r_2$)
and followed by another
real symplectic rotation (4 parameters more).
But how many of these parameters govern the two-dimensional squeezed
vacuum state? The first four parametric rotation
acting on the completely symmetric
vacuum state leaves this state unchanged.
Hence 2 squeezing parameters and 4 parameters associated with the
second rotations define the two-dimensional squeezed vacuum state.
In addition we can include an overall quantum phase factor
$e^{i\rho}$ of this state.
An explicit calculation of the general case of a two-mode squeezed
vacuum wave function is given in Ref. \cite{Schrade}.

\section{Photon statistics for the total number of photons}
We now address the question: how many of these parameters govern
the photon statistics of such a two-dimensional squeezed vacuum state?
In particular, how many of them
determine the probability of (i) counting a
total number $n$ of photons in the two modes, and (ii) counting the
number $n_1$ and $n_2$ of photons in the individual modes?

In this section we consider
the case of (i) and ask for the total number $n$ of photons in both
modes, which corresponds to the surface of a four-dimensional sphere,
$\frac{1}{2} (p_1^2+x_1^2+p_2^2+x_2^2)=n$,
centered at the origin of the four-dimensional phase space.
The rotation of the phase space does not alter this sphere
and hence only two squeezing parameters are essential.
The probability of counting the total number $n$ of photons
for the case of two independently squeezed oscillators reads
\begin{equation}
   W_n(s_1,s_2)=\sum_{n_1=0}^{n} W_{n_1}(s_1)
   W_{n-n_1}(s_2) ,
\label{eq:three}\end{equation}
where $W_{n_j}(s_j)$ is the one-dimensional
photon statistics \cite{Schleich}
\begin{equation}
   W_{n_j}(s_j)=\left\{ \begin{array}{ll} 0  & \mbox
   {for $n_j$ odd} \\
   \sqrt{1-s_j}s_j^{n_j/2}2^{-n_j}
   {n_j \choose n_j/2 } & \mbox{for $n_j$ even} \\
   \end{array} \right .
\label{eq:four}\end{equation}
of the squeezed vacuum wave function
\begin{equation}
   \Psi _{sq}(x_j) =
   \frac{e^{r_j/2}}{\pi^{1/4}} \: \exp ( -\frac{1}{2}
   \: e^{2r_j} x_j^2 )
\end{equation}
where
\begin{equation}
   s_j = \tanh^2(r_j).
\end{equation}
Equation (\ref{eq:four}) reduces Eq. (\ref{eq:three}) to
\begin{eqnarray}
   W_n(s_1,s_2) =
   \left\{ \begin{array}{ll}
   W_{2k}(s_1,s_2)  = {\cal N} 4^{-k}
   \sum\limits_{n_1=0}^{k} s_1^{n_1} s_2^{k-n_1} {2n_1 \choose n_1 }
   {2(k-n_1) \choose k-n_1 }
   \\
   W_{2k+1}(s_1,s_2) = 0
   \end{array} \right .
\label{eq:AA}\end{eqnarray}
where the normalization factor ${\cal N}$ reads
\begin{equation}
   {\cal N}=\sqrt{1-s_1}\sqrt{1-s_2}.
\end{equation}
The odd terms of $W_n$ vanish because of Eq. (\ref{eq:four}) .
The sum in Eq. (\ref{eq:AA}) has been calculated in \cite{Schrade} and the
probability of counting $n=2k$ photons then reads
\begin{equation}
   W_{2k}(s_1,s_2)={\cal N} s_2^{k}
   ~_2F_1(-k,1/2,1;1-s_1/s_2)
\label{eq:six}\end{equation}
where $_2F_1$ describes the hypergeometric function.
For the special case of identical squeezing in the two modes, that is,
$s_1=s_2=s$, Eq. (\ref{eq:six}) yields
\begin{equation}
   W_{2k}(s)={\cal N} s^k .
\label{eq:seven}\end{equation}
In Fig. 1 we show the photon statistics (\ref{eq:six})
for various magnitudes of the
squeezing parameters $s_1$ and $s_2$.
The solid and dashed curves correspond to
weak and strong symmetric squeezing
$(s_1=s_2)$,
respectively. In accordance with Eq. (\ref{eq:seven}) the photon
distribution then displays an exponential dependence.
Stronger squeezing results in a higher amount of quanta involved.
The dotted curve shows the photon statistic for an asymmetric squeezing.

\begin{center}

\plot{hpgl.1}{10.0cm}{7.cm}{}{13cm}{0.cm}{-0.5cm}{1}

\end{center}

\begin{quotation}
Fig. 1.  Probability of counting $n=2k$ photons in a
two-mode squeezed vacuum as given by Eq. (\ref{eq:six}).
For symmetric squeezing (dashed line for $s_1=s_2=0.5$ and solid line for
$s_1=s_2=0.99$) the curve is a straight line and hence exponential whereas
for asymmetric squeezing (dotted line for
$s_1=0.5$ and $s_2=0.99$ ) the photon statistics is non-exponential.
Here we have not specified the distribution of the $n=2k$ photons
among the two modes. Note that $W_{n=2k+1}=0$ which we have omitted
for simplicity.
\end{quotation}

\section{Photon statistics in the individual modes}
We now turn to the second case and
calculate the photon statistics $W(n_1,n_2)= |<n_1,n_2|\Psi_{sq}>|^2$
in the individual modes.
For this purpose we start from the generic expression
of a squeezed vacuum state $|\Psi_{sq}>$,
which we produce in the following way:
\begin{eqnarray}
   \Psi _{sq}(x_1,x_2) & & =
   e^{i \rho} \: e^{-i \phi \hat{M}} \: \hat{\Gamma}(\gamma) \:
   \hat{S_1}(r_1) \: \hat{S_2}(r_2) \: \Psi_{0,0}(x_1,x_2)
   \nonumber \\ & & \mbox{} =
   \sqrt{\frac{2}{\pi}} \:
   (A_1 B_1-C_1^2)^{1/4} \:
   e^{-Ax_1^2-Bx_2^2+2Cx_1x_2+i \rho  }
\label{eq:seven2}\end{eqnarray}
with the vacuum wave function of the two-dimensional oscillator
$ \Psi _{0,0}(x_1,x_2) = \Psi_0(x_1) \, \Psi_0(x_2) $,
the angular momentum operator
$\hat{M} = i \, (a^\dagger b - b^\dagger a)$ and the operator
$ \hat{\Gamma} (\gamma ) =
 \exp (i \gamma (b^\dagger b - a^\dagger a))$
which describes a mutual phase shift $ 2\gamma$ between the oscillators.
The wave function of a two-dimensional squeezed vacuum state
is a Gaussian described by the three complex numbers $A$, $B$, and $C$,
which are functions of the four real parameters
$r_1$, $r_2$, $\phi $, and $\gamma$.
Their explicit dependence is given in \cite{Schrade}.
One also finds there the generic two-mode squeezed vacuum wave function
depending on two more parameters that do not affect
the photon distribution.

We contract Eq. (\ref{eq:seven2}) with the
probability amplitude of the photon energy states
$\psi _{n_1}(x_1)$ and $\psi _{n_2}(x_2)$ and arrive after calculating
the resulting double integral \cite{Schrade} at
\begin{eqnarray}
   W(n_1 , n_2)  & = &
   \frac{8 \: \exp (- | \ln (n_1!/n_2!)|)}
   { |(2A+1)(2B+1)-4C^2|} \: (A_1B_1-C_1^2)^{1/2} \:
   \times  \\ & & \mbox{} \times
   \left| \frac { 4AB +2A-2B-1-4C^2 }{ 4AB -2A+2B-1-4C^2 } \right| ^
   {(n_1-n_2)/2}
   \times \nonumber \\ & & \mbox{} \times
   \left| \frac { 4AB -2A-2B+1-4C^2 }{ 4AB +2A+2B+1-4C^2 } \right| ^
   {(n_1+n_2)/2}
   \times \nonumber \\ & & \mbox{} \times
   \left| P^{|n_1-n_2|/2}_{|n_1+n_2|/2}
   \left( \frac{-4C}{\sqrt{4AB+2A+2B+1-4C^2}
   \sqrt{4C^2-4AB+2A+2B-1}} \right)  \right| ^2 .
   \nonumber
\label{result}\end{eqnarray}
Here $P_l^k$ denotes the associated Legendre Polynomial.
As in Eq. (\ref{eq:AA}) the total number of photons
$n=n_1+n_2$ must be an even number
hence $n_1$ and $n_2$ are either both even or both odd. Otherwise the
photon distribution function vanishes.

In Fig. 2 we display the probability to find $n_1$ photons in mode 1 and
$n_2$ photons in mode 2. This is the generic
case of the photon distribution
of a two-mode squeezed vacuum.
It depends on four parameters: two squeezing parameters,
the orientation of the distribution function with respect
to our laboratory system, and the correlation between the two modes.
Beside the even-odd oscillations the maxima lie
on curves which are symmetric with respect to the main diagonal $n_1=n_2$.
This behavior is similar to the distribution function of a
displaced two-mode squeezed state discussed in Ref. \cite{Caves,Artoni}.

\begin{center}

\plot{vacuum.plt}{12.0cm}{7.5cm}{}{19.5cm}{0.5cm}{0.0cm}{1}

\end{center}
\begin{quotation}
Fig. 2. Probability of counting $n_1$ and $n_2$ photons in the
two-mode squeezed vacuum calculated via Eq. (12).
We have chosen $r_1=3$, $r_2=5$, $\phi = \frac{\pi }{5}$, and
$\gamma = \frac{2 \pi}{9}$. The wavy structure of this distribution
is confined to an angle in the photon number plane.
\end{quotation}

\section{Conclusion}
We conclude by summarizing our main results.
The wave function of the two-mode squeezed vacuum depends on 6
parameters (besides the phase factor).
However only 4 of them,
-- the squeezing parameters $r_1$ and $r_2$ of the two modes,
the phase difference $\gamma$ between
the two oscillators, and the rotation of the reference system by the
angle $\phi$ --
manifest themselves in the generic case in the
distribution of the photons among the two modes.
This distribution function can be expressed explicitly in
terms of Legendre polynomials.
Only two parameters $r_1$ and $r_2$
govern the distribution of the total number of photons, which we
express explicitly in terms of a hypergeometric function.
In conclusion we want to make the remark that similar
consideration for $N$-mode squeezed vacuum state shows
that the photon distribution $W(n_1,n_2,...,n_N)$ depends
on $N^2$ parameters.

\section{Acknowledgments}
One of the authors (V.I.M.) thanks the University of Ulm and in
particular the Hans-Kupczyk Foundation for the support which made
this research possible.

\end{document}